# Low Temperature Fabrication of MgB$_2$

N. Rogado[1], M.A. Hayward[1], K.A. Regan[1], Yayu Wang[2], N.P. Ong[2],
John M. Rowell[3] and R.J. Cava[1]

[1]Department of Chemistry and Princeton Materials Institute
Princeton University, Princeton NJ

[2]Department of Physics, Princeton University, Princeton NJ

[3]Materials Research Institute, Northwestern University, Evanston IL

**Abstract**

We report the fabrication of MgB$_2$ with bulk superconducting properties by conventional solid state methods at temperatures as low as 550°C. Mg deficiencies of the type Mg$_{1-x}$B$_2$ were tested. T$_c$ was found to decrease by about 1K at large x, though the amount of non-stoichiometry, if any, is likely to be very small. For specific processing conditions, indications of the 25-30 K transition often seen in thin films were seen in the bulk materials. The lower temperature transition may be associated with the grain boundaries. These results indicate that it should be possible to fabricate MgB$_2$ with bulk properties in *in-situ* thin films at temperatures of 600°C or less.

MgB$_2$, with a T$_c$ of 39K (1), appears to offer a higher operating temperature and higher device speed than today's superconducting electronics technology based on Nb, and promises a simpler multilayer film fabrication process than HTS. For magnet and power applications, initial reports indicate, however, that the critical current density (J$_c$) is relatively low at high fields. Improved J$_c$s have been achieved in thin films by proton irradiation (2) or by the introduction of MgO (3). For electronics applications, it is highly desirable that films with T$_c$s close to 39K be made by a "single-step in-situ" process. This has not yet been achieved, although a film with T$_c$ of 23K has been made with such a process (4). The vast majority of films to date have been made by a two-step process, in which a precursor film of either B or Mg + B is annealed, typically either in Mg vapor at 900C, or in inert gas at about 600C.

The challenge in MgB$_2$ film growth is to establish the minimum deposition and growth temperature at which the film crystallizes into the compound, but at which Mg is not lost from the film, which becomes significant for temperatures above 600C (5,6). In bulk synthesis, temperatures of 900C or higher are employed to make high quality MgB$_2$ with T$_c$s near 39K and residual resistivity ratios (RRRs) of 20 or more (7). The majority of films fabricated near 600C have T$_c$s of about 25K, high resistivities, and residual resistance ratios (RRRs) close to or even less than one. Even after annealing at 900C in Mg vapor (8), many films made from precursors of Mg + B prepared by pulsed laser deposition, for example, have T$_c$s of only 25K. The origin of these non-ideal properties is not yet known with certainty, though many films contain appreciable amounts of MgO, which might be acting to both limit the MgB$_2$ grains to very small sizes and to reduce the electron mean free path within the grains. One implication of the possible presence of two superconducting energy gaps in MgB$_2$ may be that materials in the clean limit display the full 39K T$_c$, and materials in the dirty limit, as most films are believed to be, might have T$_c$s of 20-25 K (9).

To facilitate the fabrication of high quality films it is of interest to determine the minimum temperature at which MgB$_2$ can be formed with bulk properties. Here we report the fabrication of MgB$_2$ in bulk form at temperatures as low as 500 to 550°C with good superconducting properties, and hence a "single-step in-situ" film growth process should be possible at similar or even lower temperatures. This supports the single report to date (4) of film growth at 450°C, which resulted in a T$_c$ of 23K. In our bulk samples made at low temperatures, the T$_c$s in the range of 22-25K, which are commonly seen in films, were not observed. Multi-step superconducting transitions (25-30K and 38K) are observed, however, under certain conditions of processing. We attribute the lower temperature transition to weak link behavior. Finally, we present evidence for the occurrence of a small range of T$_c$s related to nominal nonstoichiometry in MgB$_2$.

Polycrystalline samples of MgB$_2$ were made by solid state reaction. Bright Mg flakes and sub-micron amorphous B powder were mixed and pressed into pellets. These pellets were placed on Ta foil in dense Al$_2$O$_3$ crucibles, and heated under a 95%Ar:5%$H_2$ gas flow. The pellets were heated, with intermediate grindings, at: (i) 550°C for 16 hours, (ii) 500°C for 2 days, and (iii) 450°C for 5 days. The samples were quenched to room temperature under the mixed Ar/$H_2$ atmosphere. Some of the resulting powders were wrapped in gold foil and hot-pressed into dense polycrystalline pellets at temperatures of 550°C under a pressure of 15 Kbar for one hour. The starting materials employed for the synthesis are very reactive. However, since bulk synthesis involves reaction over distances, which are larger than typically encountered in thin film growth, relatively long annealing times were employed to insure complete reaction. At temperatures of 900C for example, high quality $MgB_2$ is made in anneals of one hour duration (7).

Powder X-ray diffraction (XRD) with Cu K$\alpha$ radiation was employed to characterize the samples. The XRD data presented in Figure 1 show that the $MgB_2$ sample made at 550°C is single phase with no observable impurities. The XRD pattern for the sample, which was heated for 2 days at 500°C, also shown in figure 1, shows the presence of $MgB_2$ together with unreacted Mg and MgO. No $MgB_4$ is observed. The excess boron necessarily present cannot be observed by XRD. Heating this sample for longer times at 500°C did not result in any noticeable change. The sample heated at 450°C, on the other hand, did not show any observable peaks due to $MgB_2$ even after heating for 5 days.

The temperature dependent magnetization for the $MgB_2$ samples fabricated at different temperatures was measured on loose powders in a Quantum Design PPMS magnetometer with an applied DC field of 15 Oe. The zero-field cooled DC magnetization data are shown in Figure 2. $MgB_2$ synthesized at 550°C has a sharp superconducting transition, with a $T_c$ of 37 K, closely resembling the bulk properties of $MgB_2$ made at 900°C. The $MgB_2$ sample made at 500°C has a superconducting transition at 35.5 K, 1.5 degrees lower than that made at 550°C (detail in inset). It also has a smaller diamagnetic magnetization due to fact that the sample is not single phase. The sample that was heated at 450°C exhibited only very weak diamagnetism ($\cong -5 \times 10^{-3}$ emu/mol in a 15 Oe field) due to the presence of trace amounts of $MgB_2$, undetectable in the XRD patterns.

To determine whether nonstoichiometry can be the cause for the relatively low temperature superconducting transitions observed in many thin films, nonstoichiometric bulk samples were fabricated under low temperature conditions. The characterization of the superconducting transition (zero field cooled data, 15 Oe field) for Mg deficient samples fabricated at 550°C for 16 hours is shown in Figure 3. The nominal compositions are $Mg_{1-x}B_2$, with $0 \le x \le 0.7$. The systematically decreasing low temperature diamagnetic signal is consistent with a decreasing volume fraction of superconducting material with increasing Mg deficiency. XRD analysis of the samples does not reveal the presence of the compound $MgB_4$ (which would be a pure phase for $x = 0.5$) suggesting that this phase is not thermodynamically stable at low temperatures. No MgO is present. The only phase observable by XRD is $MgB_2$, indicating that the excess boron is present in the elemental amorphous state. No superconducting transitions in the 25 K range are observed. The $T_c$s are, however, lower for the most magnesium deficient samples, $x = 0.5$ and 0.7. A detail of the transition for the two limits of stoichiometry is shown in the inset. Comparison of the two XRD patterns showed a very small but discernable shift in lattice parameter for the lower $T_c$ material, and an increased diffraction peak width. The structural and/or chemical difference between these materials with different $T_c$s is therefore very subtle. Note, however, that the $T_c$s of these bulk samples, made far from stoichiometry, are still much higher than the $T_c$s of all films made at deposition/anneal temperatures of 600ºC. Hence non-stoichiometry is not likely to be the explanation for the low $T_c$s generally seen in films.

The temperature dependent resistivities of samples of polycrystalline $MgB_2$ are presented in Figure 4. The main panel shows the resistivities for two hot-pressed dense pellets. One has been heated as a powder at 900ºC for one hour and then hot-pressed at 700ºC for one hour at 15 Kbar, and the other employs the $MgB_2$ fabricated at 550ºC as described here and then hot-pressed at 550ºC for one hour at 15 Kbar. The temperature dependent resistivity for a polycrystalline sample made from the same starting materials heated at 1100ºC overnight in a sealed Ta tube is shown in the inset for comparison. The resistivity for both hot pressed samples is an order of magnitude higher than for the Ta tube prepared sample, and almost two orders of magnitude higher than observed for some $MgB_2$ samples (7). The RRRs (1.3 for both) are very similar to those observed in thin films, and are much lower than that observed for our Ta-tube prepared $MgB_2$, ~9, and high quality samples (RRR 20-30) made by that method (7). The sample hot-pressed at 700ºC displays a resistivity three times higher at all temperatures than the sample hot pressed at 550ºC. Impurities introduced during the hot-pressing process are therefore responsible for the higher resistivities, and that those impurities increase with increasing treatment temperature. XRD analysis of these samples indicates the presence of significant quantities of MgO after hot pressing, implicating oxygen as a primary factor in the degradation of the resistivity. The MgO appears to act to strongly decrease the cross section of the current path in the sample.

Finally, measurements were made at low applied DC fields in a Quantum Design MPMS SQUID magnetometer for the dense polycrystalline sample hot pressed at 550ºC. The results are shown in Figure 5. Two transitions are observed in low fields, one at approximately 28K and the other at 38K. The lower temperature transition was not seen in the powder samples before hot pressing (Figure 2). The lower temperature transition decreases in relative proportion with increasing applied field, in a manner characteristic of weak link behavior. The weak links are apparently associated with the grain boundaries. This can be seen in Figure 5 by comparison of the data taken at 5 Oe on the bulk dense polycrystalline piece, and the data for the sample piece ground into a fine powder. The low temperature transition disappears, indicating that it was associated with connections between crystallites.

Our low temperature fabrication experiments show that bulk, crystalline $MgB_2$ with $T_c$ comparable to the best polycrystalline materials can be made at temperatures of 550ºC and possibly as low as 500ºC. Given the fact that the kinetics of chemical reaction are expected to be vastly faster in thin film fabrication, this strongly suggests that high quality materials can be fabricated at low temperatures in thin films. The results also show that the low $T_c$s observed in films reported to date are not an inevitable consequence of low temperature fabrication, but are more likely due to structural or chemical defects introduced during the film fabrication process. Our results further show that the low $T_c$s observed in the films are not likely to be due to Mg nonstoichiometry induced by Mg vaporization, as, under similar conditions, suppression of $T_c$ of only a few K is observed in the bulk materials. The data suggest that oxygen may be an important harmful contaminant for $MgB_2$, but its possible influence on $MgB_2$ itself and the microstructure of bulk and thin film materials is subtle and will require further careful characterization.

**Acknowledgement**

This research was supported by grants from the U.S. National Science Foundation grant DMR – 9809483, and the Office of Naval Research, grant N00014-01-1-0920.

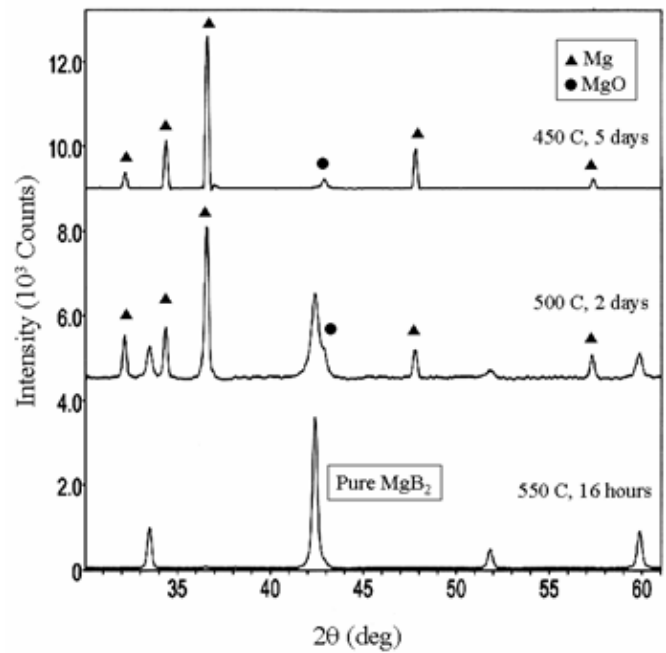

**Figure 1**. Powder X-ray diffraction patterns of MgB$_2$ samples prepared at different heating conditions. Markers are placed above the peaks corresponding to Mg and MgO impurities. The unmarked peaks belong to MgB$_2$.

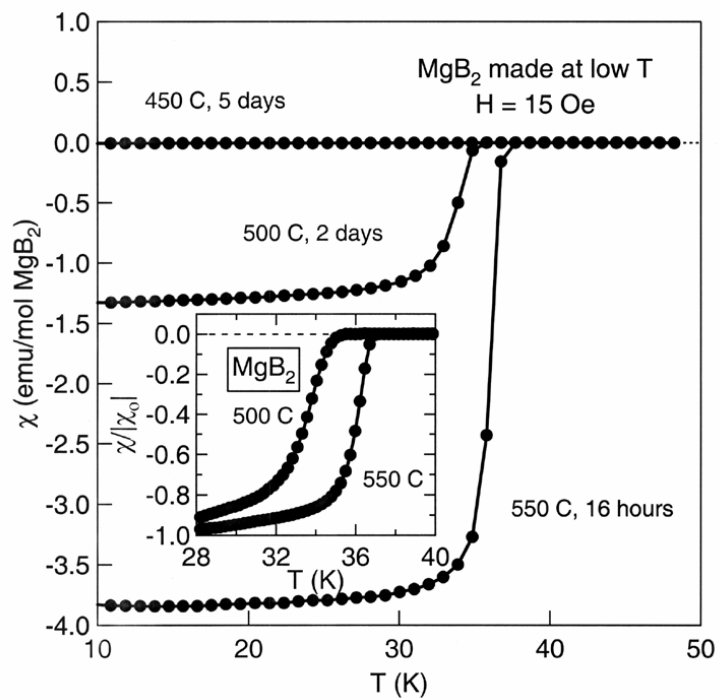

**Figure 2**. Temperature-dependent magnetization of polycrystalline powder samples of MgB$_2$ fabricated at different heating conditions. A 15 Oe DC field was applied after cooling in zero field. Inset shows detail of the region near T$_c$ for the samples prepared at 550ºC and 500ºC.

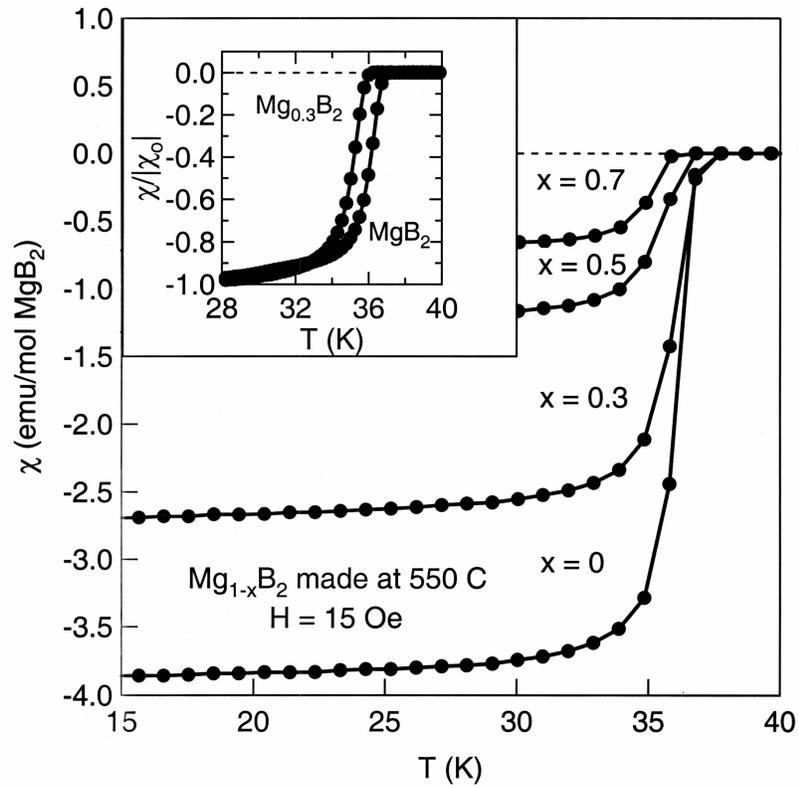

**Figure 3**. Temperature-dependent magnetization of polycrystalline powder samples of $Mg_{1-x}B_2$ fabricated at 550°C. A 15 Oe DC field was applied after cooling in zero field. Inset shows detail of the region near Tc for the samples at x = 0 and x = 0.7.

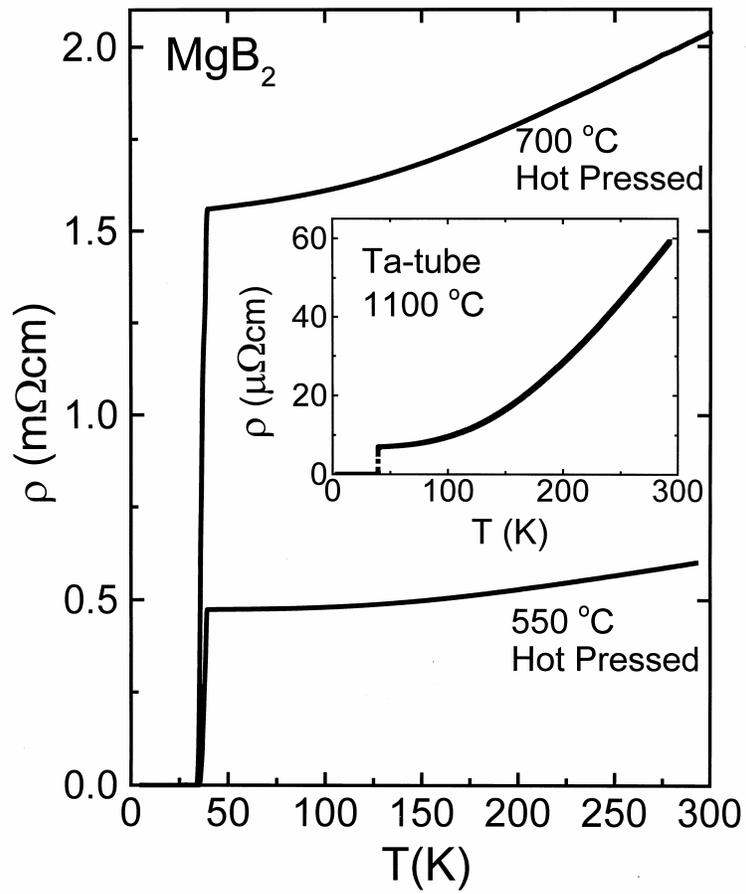

**Figure 4.** Main panel: Temperature dependent resistivities for dense polycrystalline samples hot-pressed at 700ºC and 550ºC. Inset shows temperature dependent resistivity for a polycrystalline sample made by the Ta-tube method.

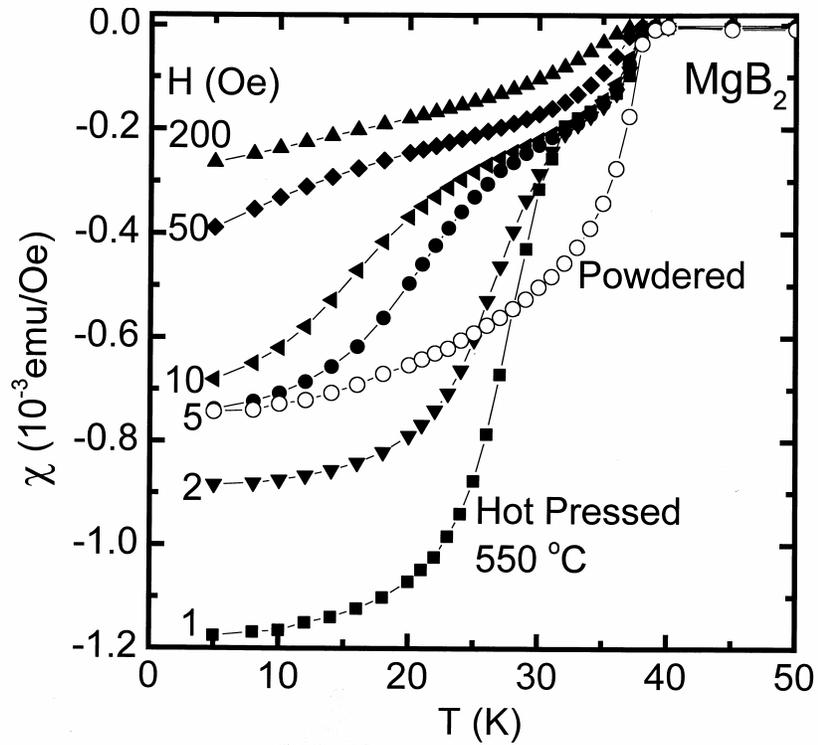

**Figure 5.** Characterization of the superconducting transition in low applied magnetic fields for a dense polycrystalline pellet of $MgB_2$ hot pressed at 550°C (closed symbols). Open symbols show the superconducting transition for the same sample pulverized into a fine powder, measured in 5 Oe applied field. The data for the powdered sample have been scaled to match the 5Oe data from the bulk piece at the lowest temperature of measurement.